%% file: vbf_v5.tex
\documentclass[preprint,1p,12pt]{elsarticle}
\usepackage{hyperref}

\usepackage{amssymb}

\journal{Physics Letters B}

\graphicspath{{Plots/}}

\DeclareGraphicsExtensions{.eps,.ps}

\input{newcommand.tex}

\def\gev{\mbox{GeV}}
\def\tev{\mbox{TeV}}
\def\ie{\emph{i.e.}}
\def\eg{\emph{e.g.}}
\def\tm{\tiny\mbox}
\def\ct{\ref}
\def\cf{\ref}
\def\as{\alpha_{\tm S}}

\begin{document}

\begin{flushright}
CERN-PH-TH/2013-086\\
CP3-13-15\\
ZU-TH 08/13
\end{flushright}
\vskip5mm


\title{Higgs production through vector-boson fusion at the NLO matched with parton showers}


\author[CERN,EPFL]{Stefano Frixione}
\author[UZH]{Paolo Torrielli}
\author[UCL]{Marco Zaro}

\address[CERN]{PH Department, TH Unit, CERN, CH-1211 Geneva 23, Switzerland}
\address[EPFL]{ITPP, EPFL, CH-1015 Lausanne, Switzerland}
\address[UZH]{Institut f\"ur Theoretische Physik, Universit\"at Z\"urich, Winterthurerstrasse 190, CH-8057 Z\"urich, Switzerland}
\address[UCL]{Centre for Cosmology, Particle Physics and Phenomenology (CP3),\\ Universit\'e Catholique de Louvain\\ Chemin du Cyclotron 2,  B-1348 Louvain-la-Neuve, Belgium}

\begin{abstract}
We present a study of Higgs hadroproduction through vector-boson fusion at the NLO in QCD matched with parton showers. We discuss the matching systematics affecting this process through a comparison of the {\amcatnlo} predictions with the {\powheg} and the pure-NLO ones.
\end{abstract}

\maketitle



\section{Introduction}

The production of a Standard-Model (SM) Higgs boson ($H^0$) through the so-called vector-boson-fusion (VBF) mechanism features the second-largest cross section among the $H^0$ production channels in hadronic collisions and, although smaller than the gluon-fusion one by about one order of magnitude, it still provides useful complementary informations. After the discovery of a SM-Higgs-like 
particle~\cite{Aad:2012tfa,Chatrchyan:2012ufa}, the emphasis is rapidly
shifting towards the determination of its properties, and in this respect VBF
may play an increasingly important role, owing to its sensitivity to 
various combinations of Higgs couplings~\cite{Plehn:2001nj}, which can be studied by considering different decay channels.  
However, the very distinctive features of VBF, with two jets lying relatively
close to the beam line and travelling in opposite directions, render it a
challenging case, given that rather severe cuts have to be applied in order to
reduce backgrounds (among which, in the coupling measurement perspective, one may count the contamination due to $gg\to
H^0$). While the typical kinematic regions probed at the LHC do not pose
problems for perturbative-QCD computations (as shown by the
behaviour of the rather moderate NLO~\cite{Han:1992hr,Figy:2003nv,
Arnold:2008rz} and NNLO~\cite{Bolzoni:2010xr,Bolzoni:2011cu} \footnote{See also \cite{Harlander:2008xn} for single-quark-line contributions.} corrections in parton-level results), the
presence of two jets in a hadronically-enriched environment implies the
necessity of using hadron-level simulations such as those generated with
parton shower Monte Carlo's (PSMC's), in order to obtain more realistic
predictions.

It has by now become a rather standard procedure that of matching NLO QCD
results with PSMC's, by using either the {\mcatnlo}~\cite{Frixione:2002ik} or
the {\powheg}~\cite{Nason:2004rx, Frixione:2007vw} formalism. Because of the
potential importance of shower and hadronisation effects and of the good
behaviour of NLO corrections, VBF appears in fact to be an ideal application
for matching techniques. However, this has been done so far only in the
context of the {\powheg} approach~\cite{Nason:2009ai}; in this letter, we
amend this by presenting {\mcatnlo} results obtained with the fully-automated
{\amcatnlo} framework, and by comparing them extensively with those 
obtained with the code constructed in reference \cite{Nason:2009ai} and implemented in the publicly available {\powheg}-{\textsc{Box}} framework \cite{Alioli:2010xd}.
The primary motivation for doing so is phenomenological. As is known,
{\mcatnlo} and {\powheg} differ by terms of order ${\cal O}(\as^{b+2})$~\cite{Nason:2012pr},
\ie~two orders larger than the Born's; furthermore, they differ by logarithmic
orders beyond the leading even if matched to the same PSMC, owing to the
fact that {\powheg} generates the first emission with own Sudakov form
factors, independent of those of the PSMC\footnote{The latter differences are
actually logarithmically leading in the case of an angular-ordered
PSMC which does not include a vetoed-truncated shower \cite{Nason:2004rx}.}.
While these differences are typically small, consistently with their being beyond the nominal accuracy of the calculations, Higgs production in gluon fusion constitutes a striking counter-example, with the two approaches yielding significant discrepancies
in the Higgs transverse momentum\footnote{Before any tuning of the {\tt hfact} parameter in \powheg.}, and in the Higgs--hardest-jet rapidity
difference. The latter observable in particular, being quite sensitive to
the radiation pattern generated by the PSMC\footnote{See
reference \cite{Frederix:2012ps} for a discussion on this point.}, plus
the internal Sudakov in the case of \powheg, could have direct implications
for VBF, given the importance of 'extra' radiation in this process. In
general, the differences between the {\mcatnlo} and {\powheg} results should
give one a fair idea of the NLO-matching systematics, a topic which, to the
best of our knowledge, has not been studied in VBF Higgs production. A
lesser motivation is technical, and is that of validating the {\amcatnlo}
machinery with a further non-trivial process on top of those considered so
far.

We remind the reader that {\amcatnlo} is a generator that implements the
matching of a generic NLO QCD computation with a PSMC according to the
{\mcatnlo} formalism; its defining feature is that all ingredients of such
matching and computation are fully automated. The program is developed within
the \madgraph\,5 \cite{Alwall:2011uj} framework and, as such, it does not
necessitate of any coding by the user, the specification of the process and of
its basic physics features (\eg~particle masses or phase-space cuts) being
the only external informations required: the relevant computer codes are then
generated on-the-fly, and the only practical limitation is represented by CPU
availability.
{\amcatnlo} is based on different building blocks, each devoted to the
generation and evaluation of a specific contribution to an NLO-matched
computation. \madfks~\cite{Frederix:2009yq} deals with the Born and
real-emission terms, and in particular it performs, according to the FKS prescription
\cite{Frixione:1995ms,Frixione:1997np}, the
subtraction of the infrared singularities that appear in the latter
matrix elements; moreover, it is also responsible for
the process-independent generation of the so-called Monte Carlo subtraction
terms, namely the contributions that prevent any double-counting in
the {\mcatnlo} cross sections. Finally, \madloop~\cite{Hirschi:2011pa} 
computes the finite part of the virtual contributions, using the OPP
\cite{Ossola:2006us} one-loop integrand-reduction method and its
implementation in \cuttools~\cite{Ossola:2007ax}.

\section{Results}
\label{secamc:vbf}
In this section we present results relevant to the production of a $125~\gev$
Standard-Model Higgs boson through a VBF mechanism at the $8~\tev$ LHC.  {\amcatnlo} includes all interferences between $t$- and $u$-channel diagrams,
such as those occurring for same-flavour quark scattering and for partonic
channels that can be obtained by the exchange of either a $Z^0$ or a $W^\pm$
boson (\eg~$ud\to H^0 ud$). These interferences, which are kinematically suppressed and {\it de facto} negligible, are not included in \powheg. Furthermore, only vertex loop-corrections are considered in both matching schemes, as the omitted loops are totally negligible \cite{Ciccolini:2007ec}. Electroweak NLO corrections, negative and of the order of 5\% for this Higgs mass and collider energy \cite{Ciccolini:2007ec,Ciccolini:2007jr}, are not included. The Higgs boson is considered as stable.

Matching with different showers, namely {\herwigsix} \cite{Corcella:2000bw}, {\herwigpp} \cite{Bahr:2008pv}, and
virtuality-ordered {\pythiasix} \cite{Sjostrand:2006za} (abbreviated in the following with HW6, HWPP,
and PY6, respectively), is considered both in {\amcatnlo} and in
{\powheg}, in order to estimate the dependence of physics results on the
shower model, within the same matching scheme.\\

\subsection{Setup: parameters and cuts}
\label{secamc:params}
Here we list the input settings employed in this computation.  The values for
the Standard-Model parameters follow the prescriptions of the Higgs
Cross-Section Working Group (HXSWG)~\cite{hxswgwiki}:
\begin{eqnarray}
    &&M_W = 80.398~\gev\,,\qquad\qquad \Gamma_W = 2.089~\gev\,, \nonumber\\
    &&M_Z = 91.188~\gev\,,\qquad\qquad~ \Gamma_Z = 2.496~\gev\,, \nonumber\\
    &&G_F = 1.166 \times 10^{-5}~\gev^{-2}\,.
\end{eqnarray}
Results are obtained by using the MSTW2008NLO PDF set~\cite{Martin:2009iq},
with errors estimated at the $68\%$ confidence level. Moreover,
renormalisation and factorisation scales are set equal to the $W$ mass, as
suggested by the HXSWG. All parton showers are run with their default settings, with the only exception of \amcatnlo+PY6, where \texttt{PARP(67)} and \texttt{PARP(71)} are set equal to one. Furthermore, no simulation of the underlying event is
performed.

Parton-level events are generated without imposing generation cuts, with the exception of a technical cut that requires at least two jets with $p_T(j) >
2~\gev$ in the {\amcatnlo} samples. This cut has been extensively checked not
to introduce any bias in total rates and differential distributions. After
shower and hadronisation, typical selection cuts used in experimental VBF
analyses (called \emph{VBF cuts} henceforth) are applied: hadrons are
clustered into jets by using the anti-$k_T$ algorithm~\cite{Cacciari:2008gp} as
implemented in {\fastjet}~\cite{Cacciari:2011ma}, with $\Delta R = 0.5$. The
presence of at least two jets is required, with $p_T(j) > 20~\gev$ and $|y(j)|<
4.5$. Furthermore, the two hardest jets (\ie~the two jets with the largest
transverse momenta) among those fulfilling these criteria are required to
have an invariant mass $M(j_1,j_2)>600~\gev$ and a rapidity separation
$\left|\Delta y(j_1,j_2)\right| > 4$.


\subsection{Differential distributions}
\label{secamc:results}
We now present results for various differential distributions. The same
pattern is adopted for all figures. In the main frame of each figure, the three
curves that correspond to the {\amcatnlo} samples are shown: black solid for
HW6, red dashed for PY6, and blue dot-dashed for HWPP. The upper and central
insets show, with the same colours and patterns as the main frame, the ratios
of the {\amcatnlo} and {\powheg} results over the fixed-order
NLO ones\footnote{Also computed with {\amcatnlo} and cross-checked against existing
  results.}, in order to assess the impact of the different parton showers and
matching schemes on the observables considered. The lower insets show the
scale (red dashed) and PDF (black solid) uncertainties relevant to the
{\amcatnlo}+HW6 sample. The scale-variation band is the envelope of the
results obtained by varying independently the factorisation and
renormalisation scales in the ranges
\begin{equation}
    \frac{M_W}{2} < \mu_R , \mu_F < 2M_W\,,
\end{equation}
while the PDF errors are computed with the Hessian method~\cite{Martin:2002aw}, as prescribed by the MSTW set. We remind the reader that the {\amcatnlo} Les Houches parton-level event files store additional information sufficient to the automatic determination of scale and PDF uncertainties at no extra CPU cost, by means of the reweighting technique presented in \cite{Frederix:2011ss}.

Part of the differences that will appear in the upper and middle insets is due
to the different impact of the VBF cuts on the QCD radiation generated by the
various PSMC's. To better understand this effect, in table \ct{tabamc:ratios}
we quote the ratios of the matched-NLO cross section after VBF cuts to the fixed-order NLO one, $\sigma^{\tm{NLO}}_{\tm{CUTS}}=0.388(2)\pb$. It can be
highlighted that these ratios are all smaller than one, as typically parton
showers tend to spread the radiation hardness throughout the phase space, causing slightly more events to fail the cuts. On top of this, there is a
clear pattern $\sigma^{\tm{HW6}}_{\tm{CUTS}} > \sigma^{\tm{PY6}}_{\tm{CUTS}} > \sigma^{\tm{HWPP}}_{\tm{CUTS}}$
both for {\amcatnlo} and {\powheg}.

\begin{table}
    \centering
    \begin{tabular}{|c|c|c|c|}
        \hline
        & $ \herwigsix $ & \pythiasix & \herwigpp \\
        \hline
        \amcatnlo & $\qquad 0.93 \qquad $ & $\qquad 0.89 \qquad$ & $\qquad 0.83 \qquad$ \\ 
        \powheg & $0.92$ & $0.86$ & $0.83$ \\ 
        \hline
    \end{tabular}
    \caption{\footnotesize
    Ratios of matched-NLO cross sections to the fixed-order NLO one after VBF cuts for {\amcatnlo} and {\powheg}.
    \label{tabamc:ratios}
    }
\end{table}
\begin{figure}[ht!]
\centering
\includegraphics[width=0.75\textwidth]{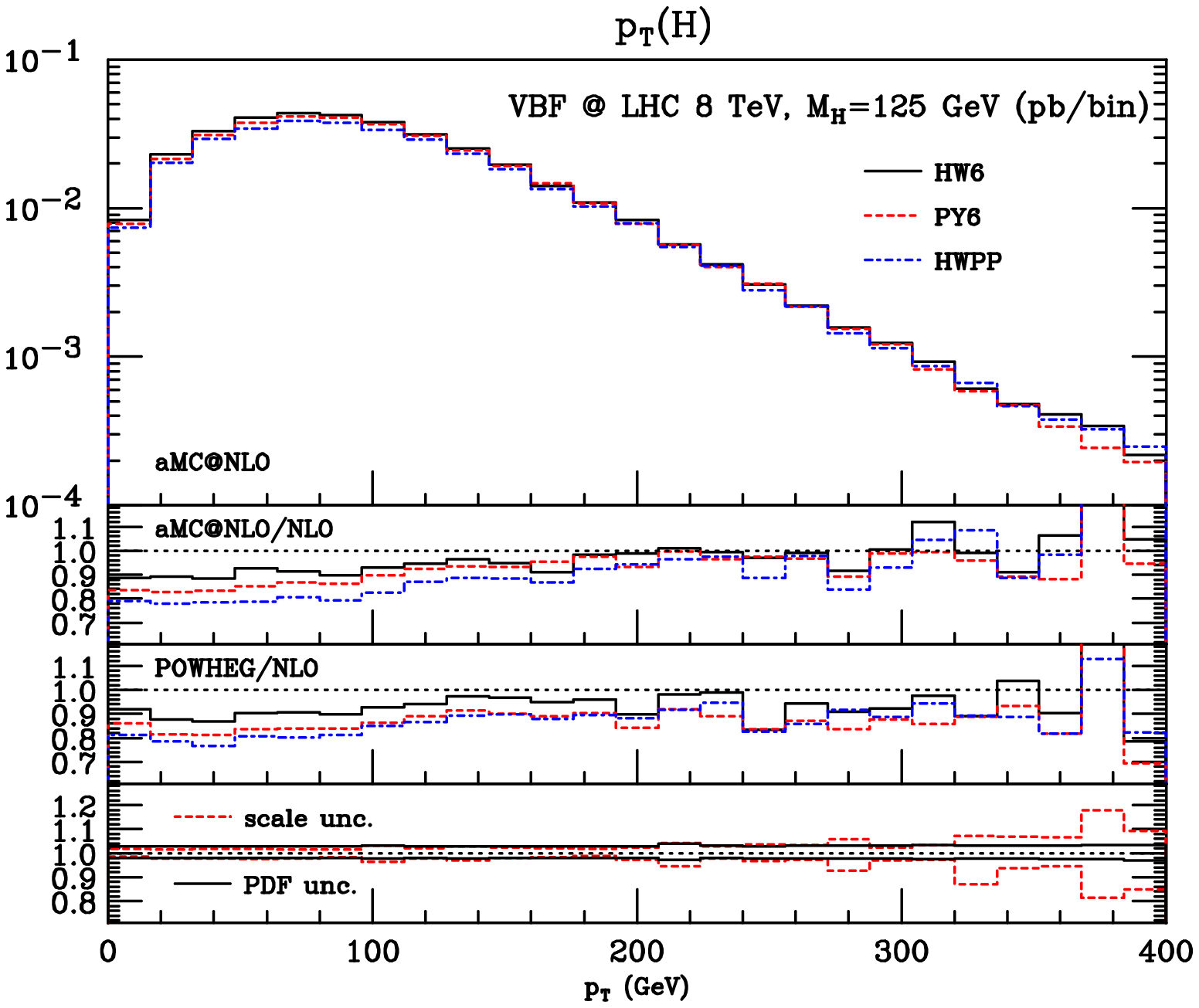}\\[5mm]
\includegraphics[width=0.75\textwidth]{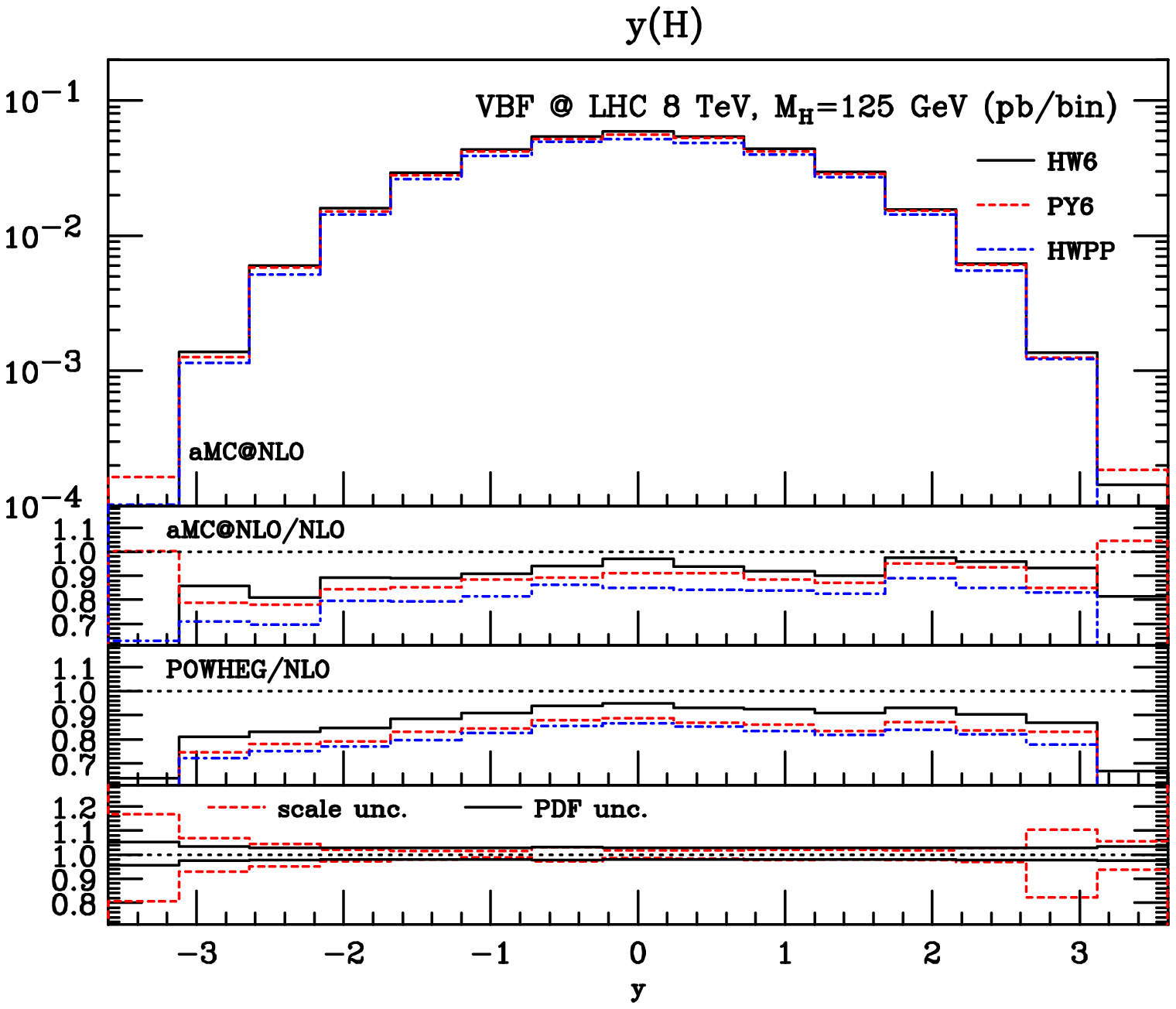}
\caption{\footnotesize
  \label{figamc:higgs}
  Higgs boson transverse-momentum (top) and rapidity (bottom) distributions. \captionamcatnlo}
\end{figure}
\begin{figure}[ht!]
\centering
\includegraphics[width=0.75\textwidth]{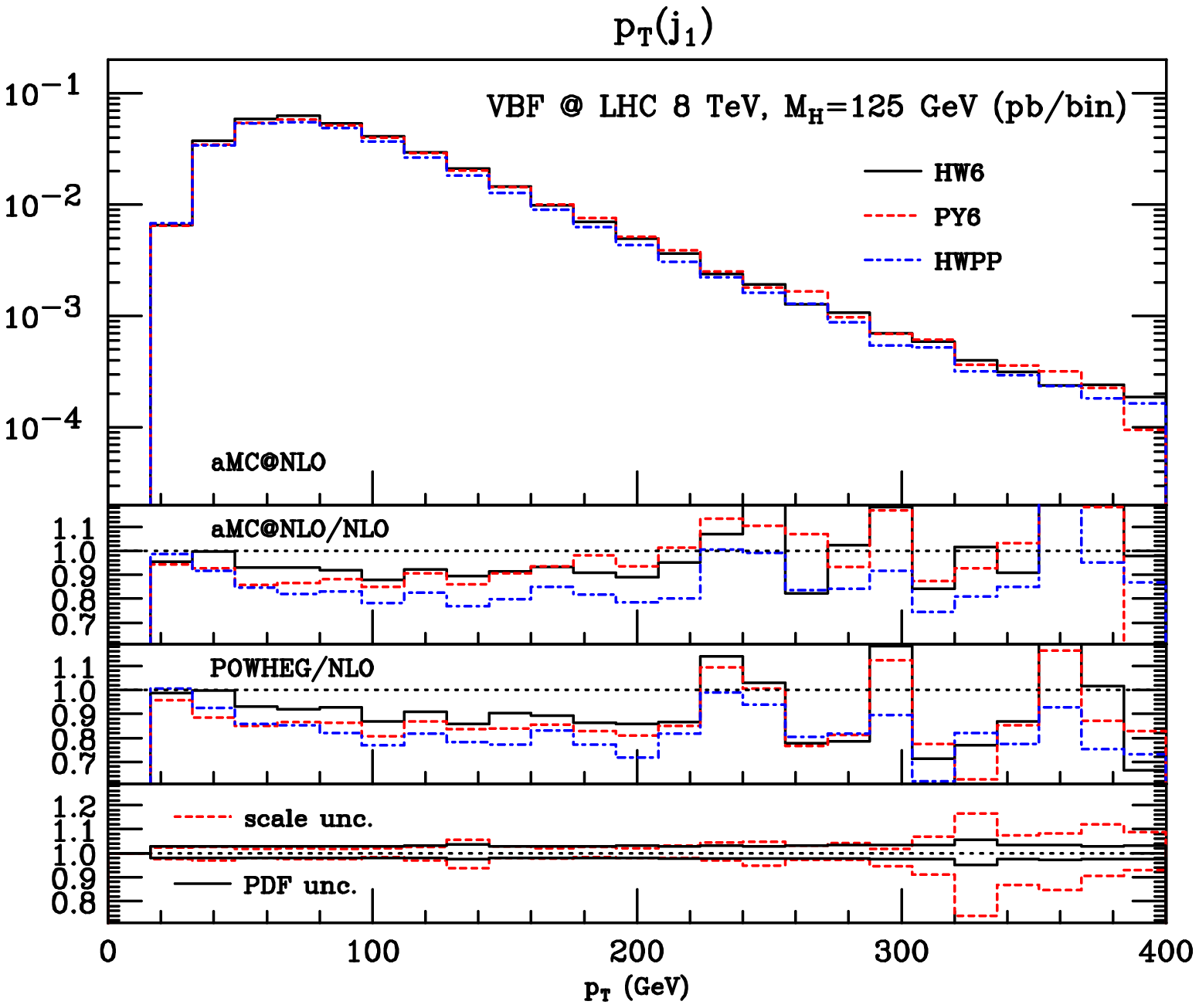}\\[5mm]
\includegraphics[width=0.75\textwidth]{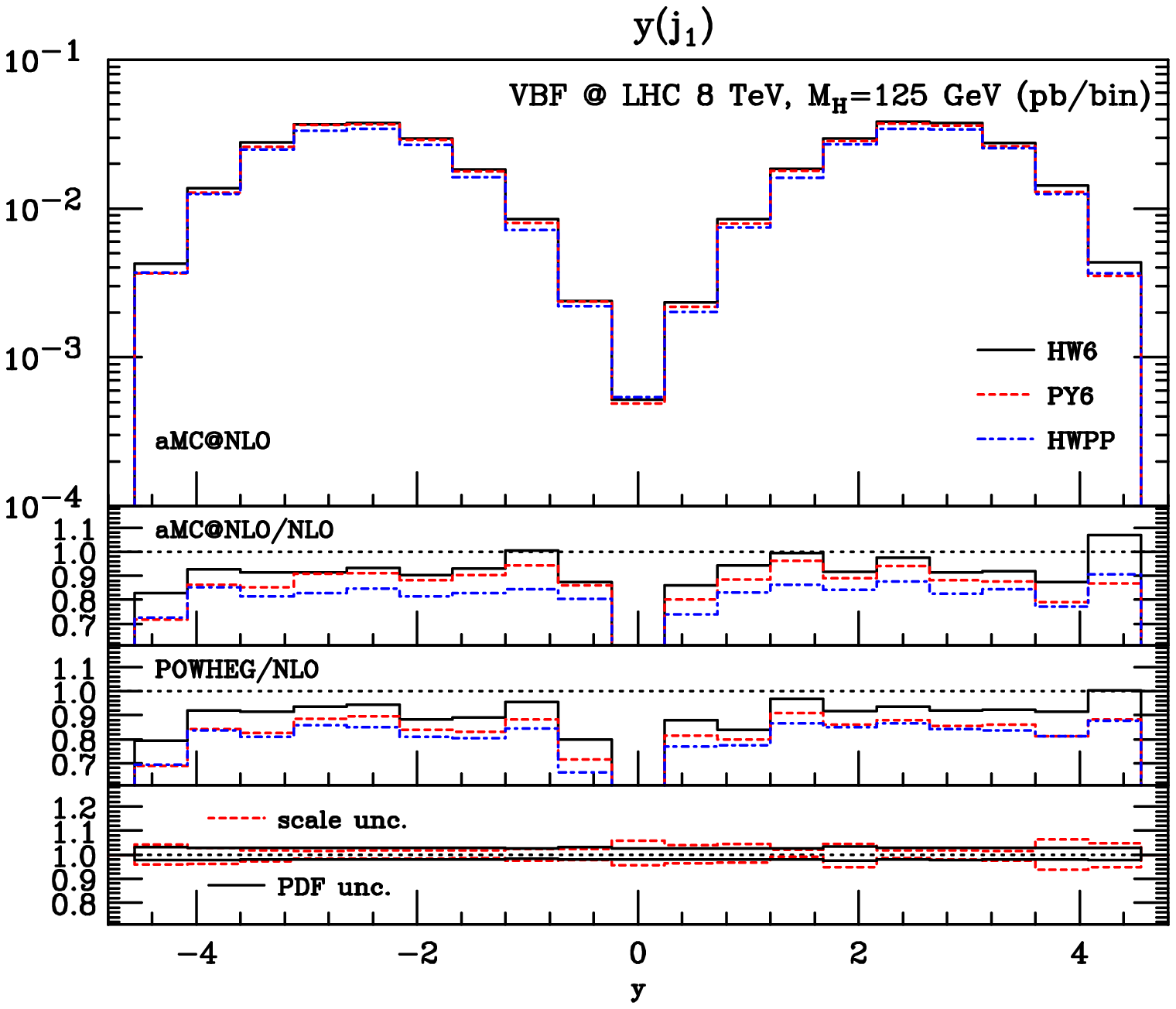}
\caption{\footnotesize
  \label{figamc:j1}
  Same pattern as in figure \ref{figamc:higgs} for the hardest-jet transverse momentum (top) and rapidity (bottom).}
\end{figure}
\begin{figure}[ht!]
\centering
\includegraphics[width=0.75\textwidth]{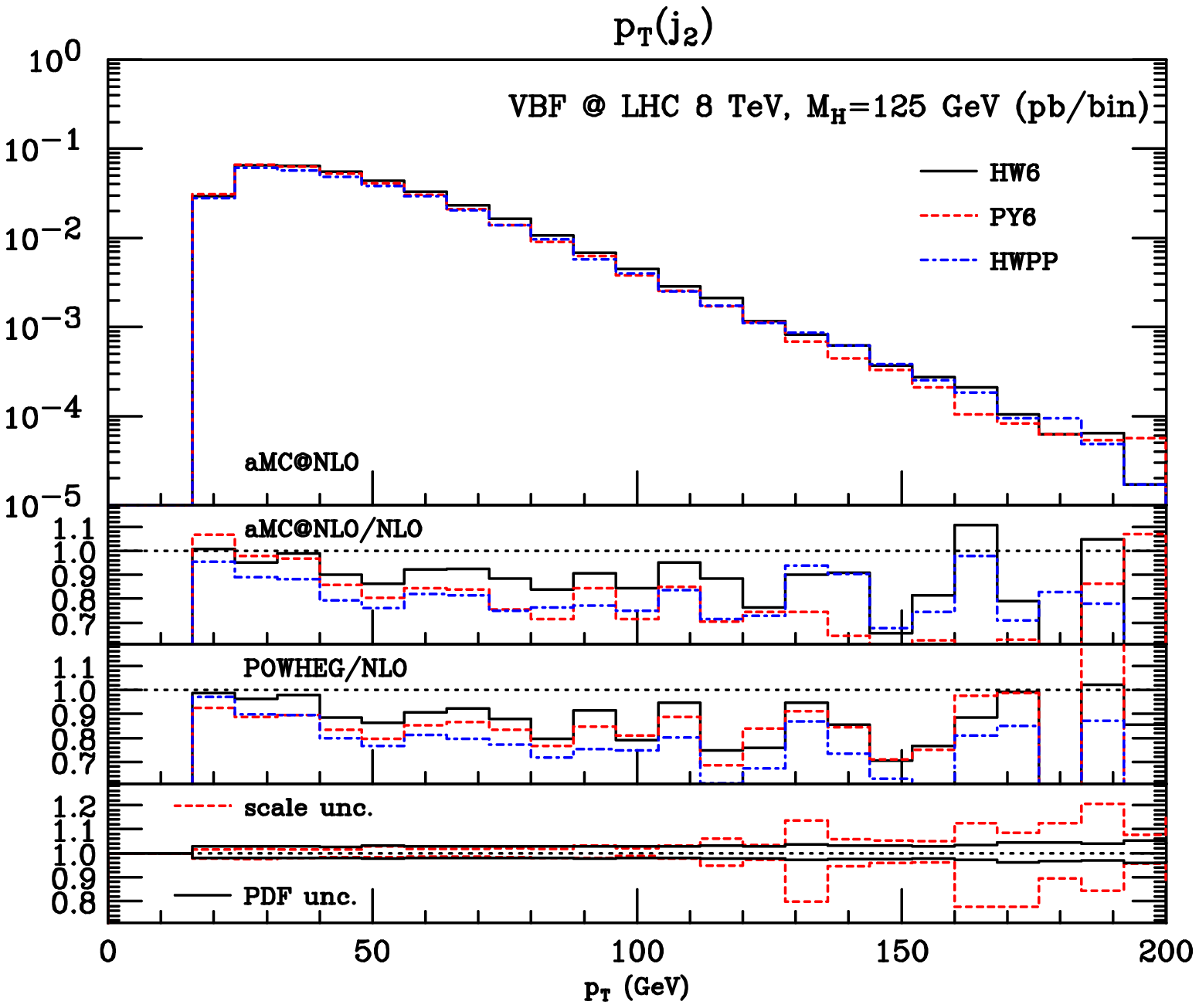}\\[5mm]
\includegraphics[width=0.75\textwidth]{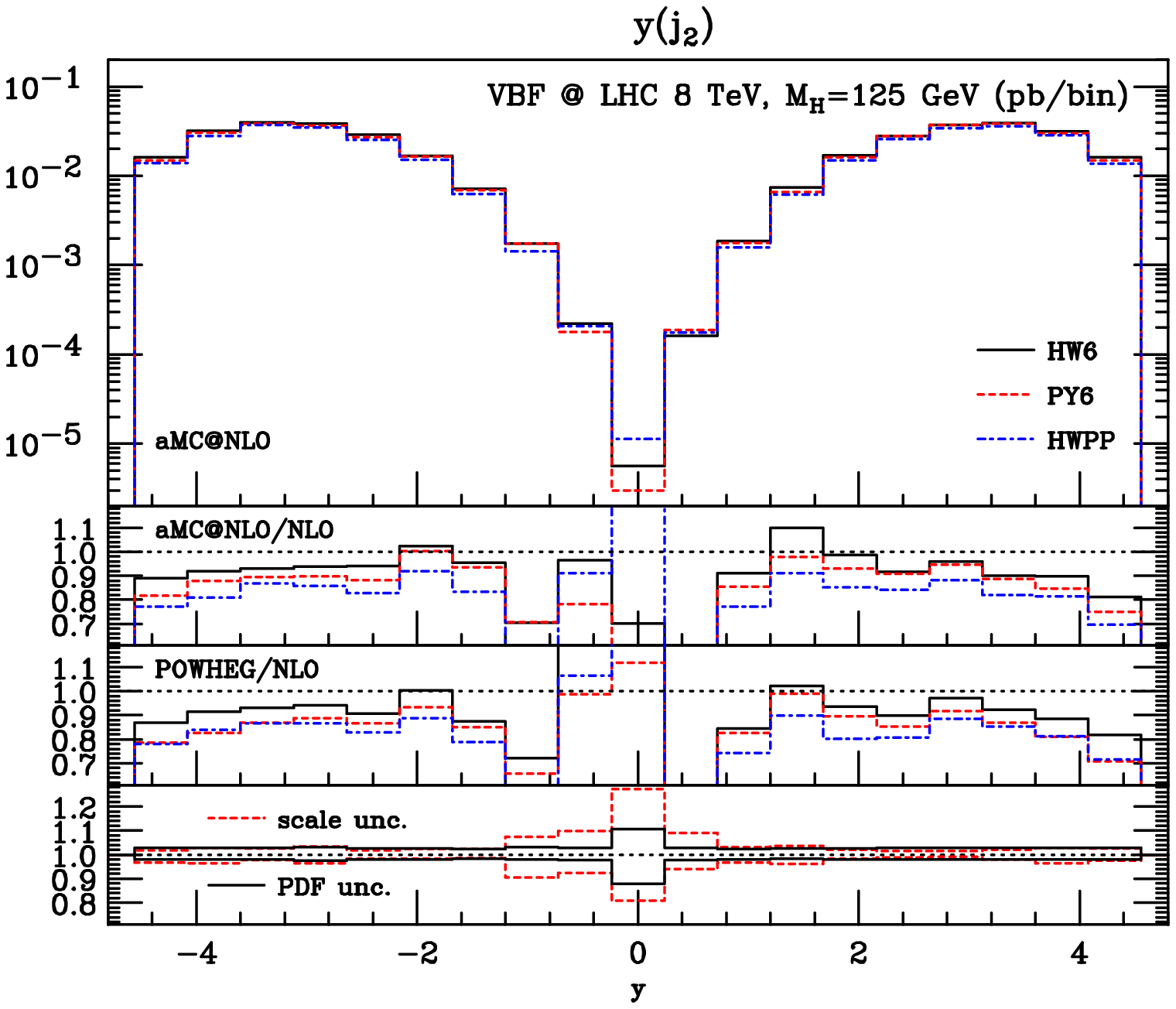}
\caption{\footnotesize
  \label{figamc:j2}
    Same pattern as in figure \ref{figamc:higgs} for the second hardest-jet transverse momentum (top) and rapidity (bottom).}
\end{figure}
In figures \cf{figamc:higgs} to \cf{figamc:j2} we show the transverse momentum
and rapidity of the Higgs boson and of the two hardest (tagging) jets. All these observables are described with NLO accuracy, as they are non-trivial in their full kinematic ranges already at the Born level $\mathcal O(\as^0)$. Therefore, general agreement among the two different matching frameworks, as well as among different showers is expected. Indeed, all NLO-matched curves are fairly compatible with each other once the ratios in table \ct{tabamc:ratios} and the theoretical uncertainties in the lower insets are taken into account. The comparison with the fixed-order NLO prediction, on top of the overall normalisation effect already shown in table~\ref{tabamc:ratios}, displays a consistent action of the shower in affecting the jet spectra, an effect which is increasingly important as one moves downwards in the jet hierarchy (\ie~from the hardest to the softest jets); in fact this trend will become even more evident in the case of the third jet (see later). As a consequence of the recoil against the shower-enriched jet activity, NLO-matched curves display harder and more central Higgs-boson distributions.
For the observables shown in these figures, PDF and scale uncertainties are generally small (typically of the order of $\pm$3\% to $\pm$5\%), and fairly constant, with only mild increases at large transverse momenta.

Similar conclusions as the ones presented above can be drawn for the azimuthal
separation between the two tagging jets, displayed in the top plot of figure
\cf{figamc:dphi12-n}, which also shows excellent shape agreement between fixed-order and matched computations. This is reassuring, since this observable is
particularly sensitive to Higgs-boson quantum numbers as spin and parity~\cite{Hagiwara:2009wt,Englert:2012xt,Djouadi:2013yb}, and
therefore any theoretical uncertainty is reflected on the characterisation of
Higgs properties.

\begin{figure}[ht!]
\centering
\includegraphics[width=0.75\textwidth]{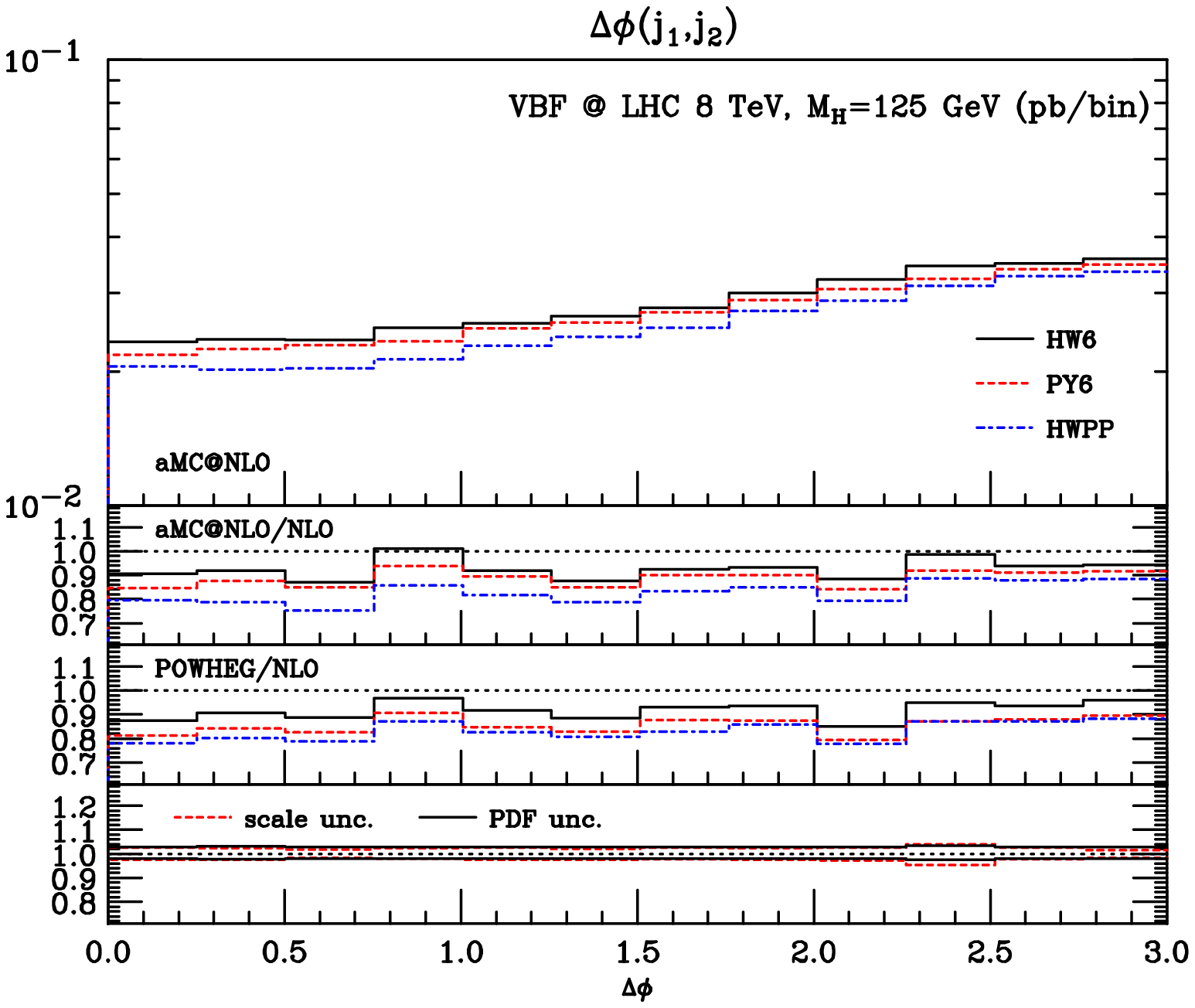}\\[5mm]
\includegraphics[width=0.75\textwidth]{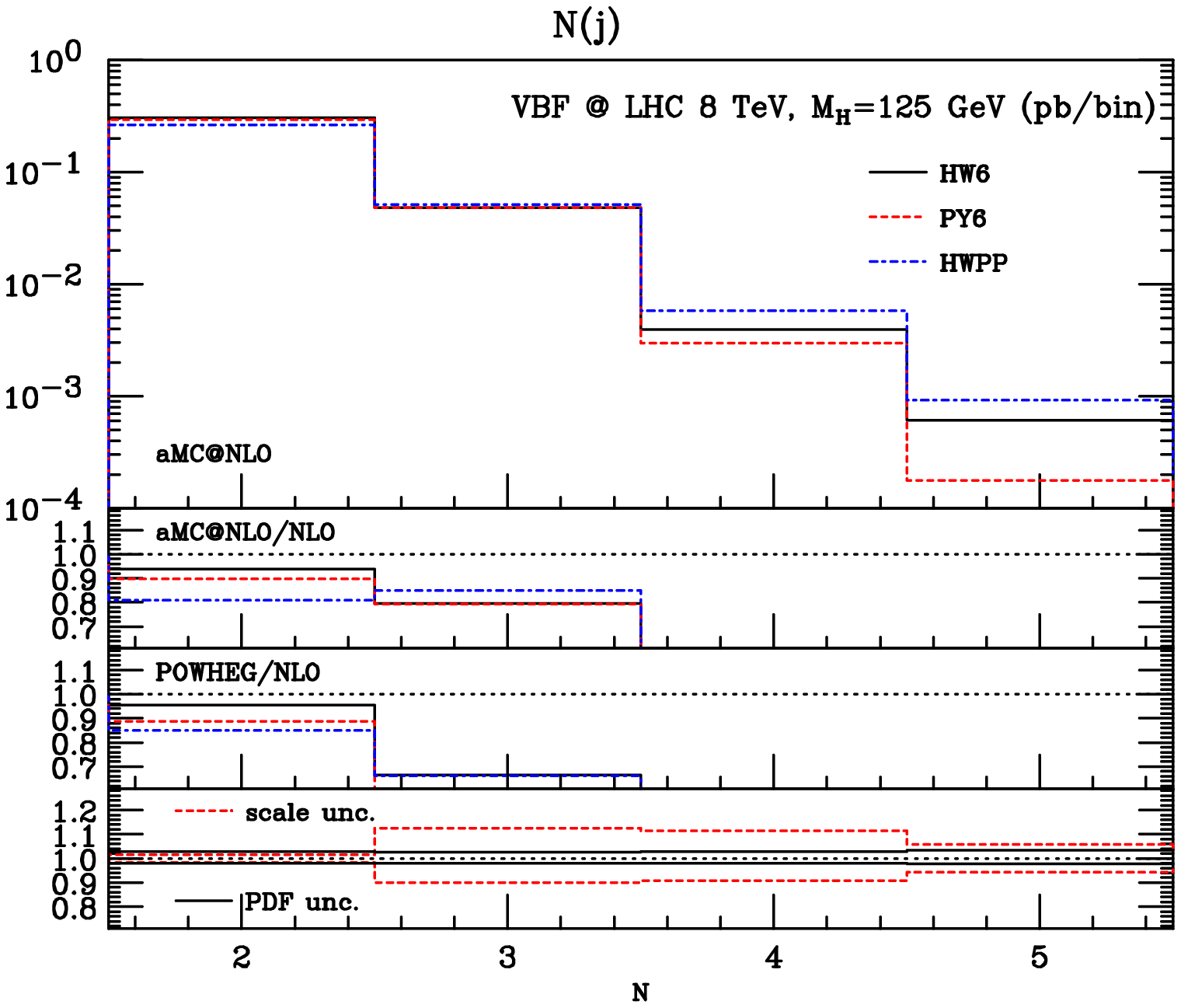}
\caption{\footnotesize
  \label{figamc:dphi12-n}
    Same pattern as in figure \ref{figamc:higgs} for the azimuthal separation of the two hardest jets (top) and the exclusive jet-multiplicities (bottom).}
\end{figure}
\begin{figure}[ht!]
\centering
\includegraphics[width=0.75\textwidth]{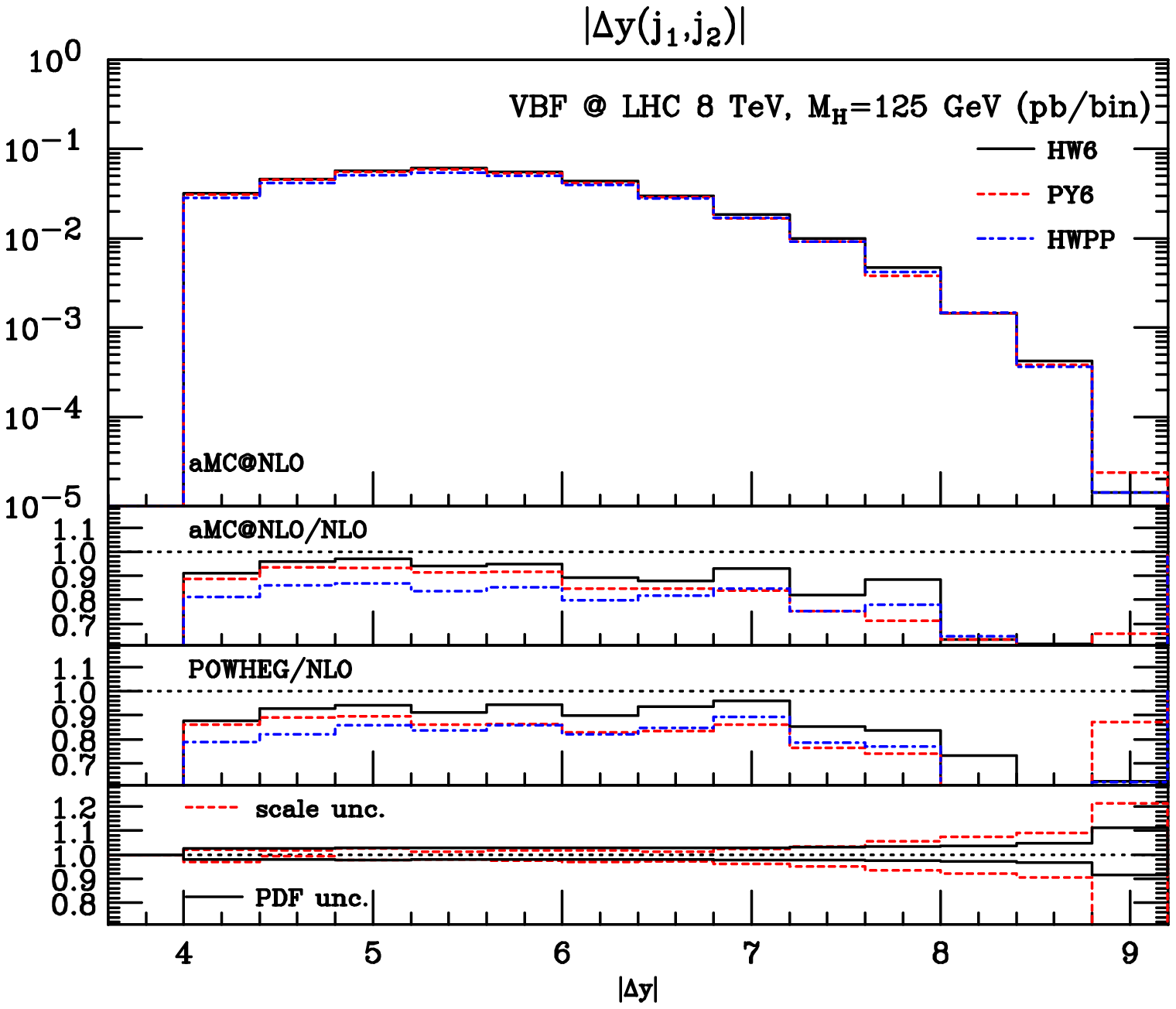}\\[5mm]
\includegraphics[width=0.75\textwidth]{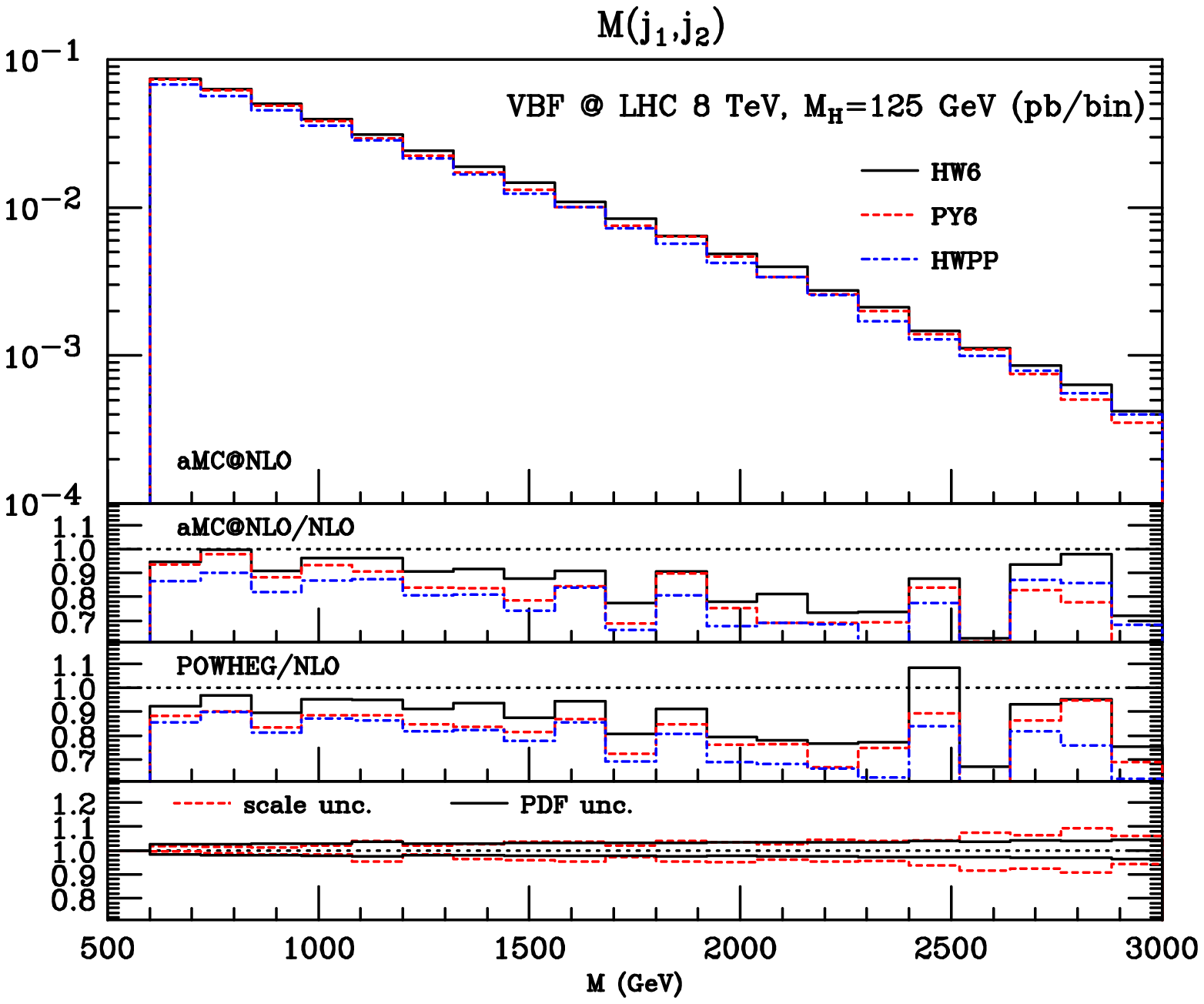}
\caption{\footnotesize
\label{figamc:dym12}
     Same pattern as in figure \ref{figamc:higgs} for the rapidity separation (top) and invariant mass (bottom) of the two hardest jets.}
\end{figure}
Showering effects are more important for observables such as the invariant
mass and the rapidity separation of the two hardest jets. These quantities,
shown in figure \cf{figamc:dym12}, can probe extreme kinematic configurations,
where the two jets lie at very large rapidities in opposite hemispheres. At
fixed-order and at the parton level, events with large invariant mass and
rapidity separation involve partons with energies of up to $\mathcal
O(1~\tev)$.  Such energetic partons, when processed through the shower,
have a large probability to fragment several times, so that the resulting
hadron-level jets may carry only a small fraction of their energy. This
results in the 20\%- to 30\%-deficit, with respect to the fixed-order NLO,
visible in the matched curves at the rightmost edge of these distributions. In
the same region, the theoretical uncertainties grow up to the level of $\pm
10$-$15\%$, especially at large rapidity difference. For the PDF's this
corresponds to the larger uncertainty at $x\sim 1 $, whereas for the scale
uncertainty this can be understood as the inadequacy of the choice
$\mu_{R,F} = M_W$ for such extreme kinematics. Indeed we have checked that, by
employing a dynamical scale like $\mu_{R,F}=\sum_i p_{T,i}/2$ (the sum running
over final-state partons), the increase in the scale-uncertainty band at large rapidity
difference is very much reduced, to the level of $\pm$5\%, with a negligible
shift in the central value.

Observables relevant to the third-hardest jet are more sensitive to the
different matching procedures and to the effects of parton showers, since their
description at the matrix-element level is only LO. In the lower plot of
figure \cf{figamc:dphi12-n} the exclusive jet-multiplicity is shown. While the
2-jet bin closely follows the ratios in table \ct{tabamc:ratios}, the 3-jet
bin shows larger differences, with {\powheg} predicting less events than
\amcatnlo. The deficit with respect to the fixed-order result in this bin ranges from 15\% to 20\% for {\amcatnlo}, whereas it is larger, 30\% or more, for
{\powheg}, irrespectively of the PSMC employed. This indicates that such an 
effect is mainly a matching systematics rather than being induced by different
shower models. It has to be stressed, however, that the {\powheg} and
{\amcatnlo} predictions are still compatible within scale uncertainties, which
for the 3-jet bin are about $\pm$10\%, consistently with the LO precision of
this observable. From the 4-jet bin onwards, the description is completely driven by the leading-logarithmic (LL) accuracy of the showers and by the tunes employed,
which translates in large differences among the various generators. For such
jet multiplicities, theoretical-uncertainty bands are completely
unrepresentative.

The 3-jet-bin pattern described above determines the normalisation of the
third-jet transverse-momentum and rapidity distributions, shown in figure
\cf{figamc:j3}. These variables can be significantly affected by the different
radiation produced by the PSMC's. In particular the {\amcatnlo}
results, which are quite close to the pure NLO on average, display a $\pm$15\%
dependence on the shower adopted. Conversely, the three {\powheg} predictions
are slightly closer to each other; we reckon that this is a consequence of 
the hardest emission being generated in this formalism by a Sudakov
which is independent of the actual PSMC employed, as already mentioned
in the introduction. The {\powheg} curves show a 30\%-deficit with respect to the pure NLO, and are more central and softer than the {\amcatnlo} ones. Different settings in
\pythiasix~have been checked to induce a variation in the results within the
previously mentioned discrepancy range, which is thus to be considered as a
genuine measure of the matching systematics affecting these quantities. Scale
uncertainties are again compatible with the LO nature of these observables, of the order of $\pm$10\% in the whole rapidity range, and growing from $\pm$10\% to $\pm$20\% with transverse momentum.

\begin{figure}[ht!]
\centering
\includegraphics[width=0.75\textwidth]{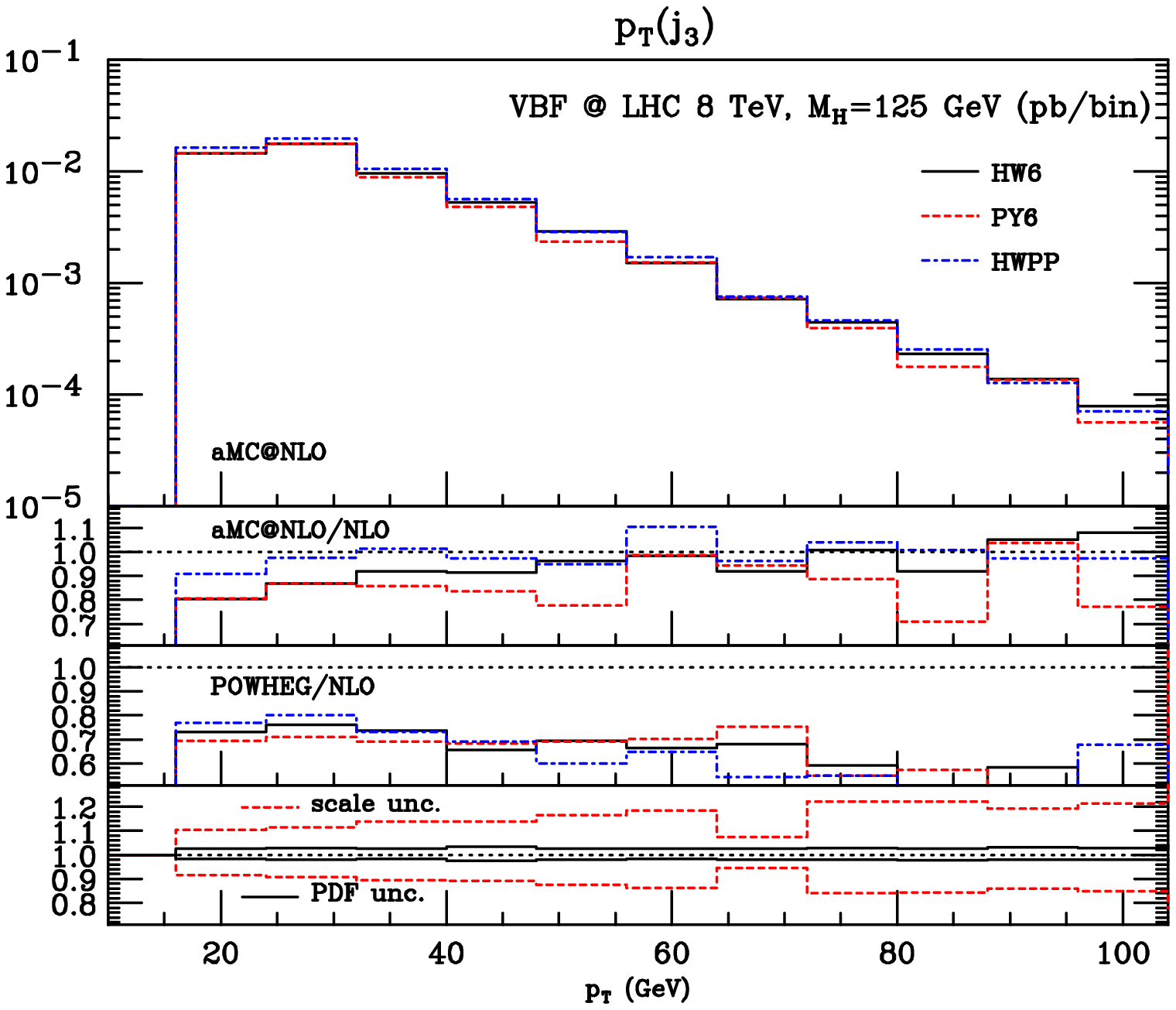}\\[5mm]
\includegraphics[width=0.75\textwidth]{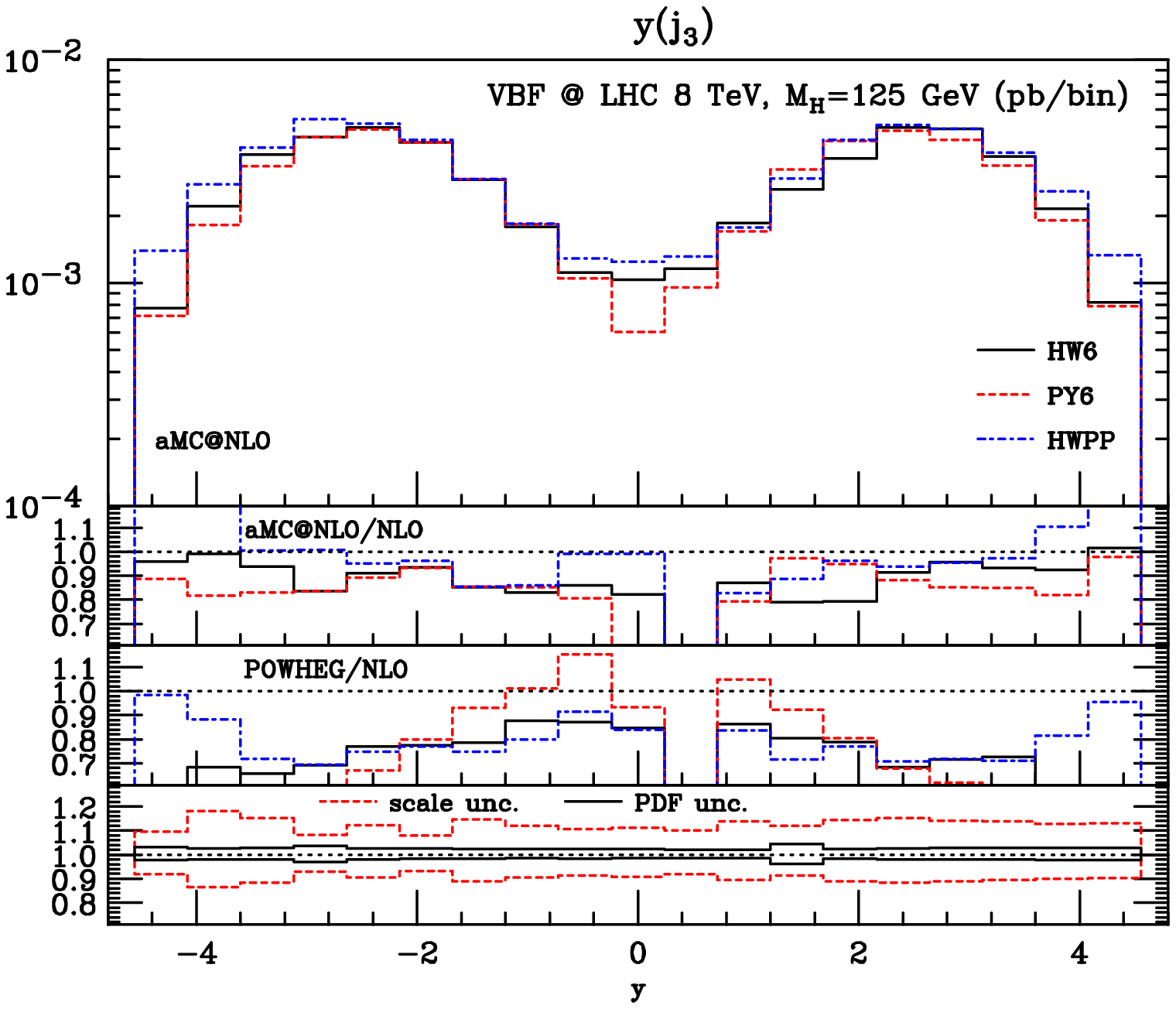}
\caption{\footnotesize
  \label{figamc:j3}
    Same pattern as in figure \ref{figamc:higgs} for the third hardest-jet transverse momentum (bottom) and rapidity (top).}
\end{figure}
Because of the peculiar radiation pattern of VBF, which favours QCD emissions
far from the central-rapidity region, one way to reduce the contamination
due to background processes, as well as from other production channels
(\eg~gluon fusion), is that of rejecting any event featuring a \emph{veto jet}, 
namely an extra jet with rapidity lying between those of the two hardest jets:
\begin{equation}
    \min\{y(j_1),y(j_2)\} < y(j_{\tm{veto}}) < \max\{y(j_1),y(j_2)\}\,.
    \label{eq:veto}
\end{equation}
The predictions for the transverse momentum and rapidity of the veto jet are
shown in figure \cf{figamc:jveto}.  The definition in equation (\ref{eq:veto})
implies that the more central the third jet, the larger the probability that
it be the veto jet. Since {\powheg} predicts a more central third jet with
respect to {\amcatnlo}, the veto condition has the effect that the two
predictions for the veto jet are slightly closer to each other than for the
third jet. {\amcatnlo} yields visibly softer and less central distributions, with discrepancies of 20\% to 30\% with respect to the pure NLO at small transverse momentum and large rapidity. The {\powheg} predictions for the transverse momentum are softer than {\amcatnlo} (with the exception of the matching to PY6, where shapes are similar), while rapidities are more central, with 20\%- to 30\%-discrepancies with respect to the pure NLO at the edges of the spectra.
As was the case for the third-hardest jet, the observables  related to the veto jet are described only at LO accuracy, and affected by large uncertainties, roughly $\pm$15\%.
\begin{figure}[ht!]
\centering
\includegraphics[width=0.75\textwidth]{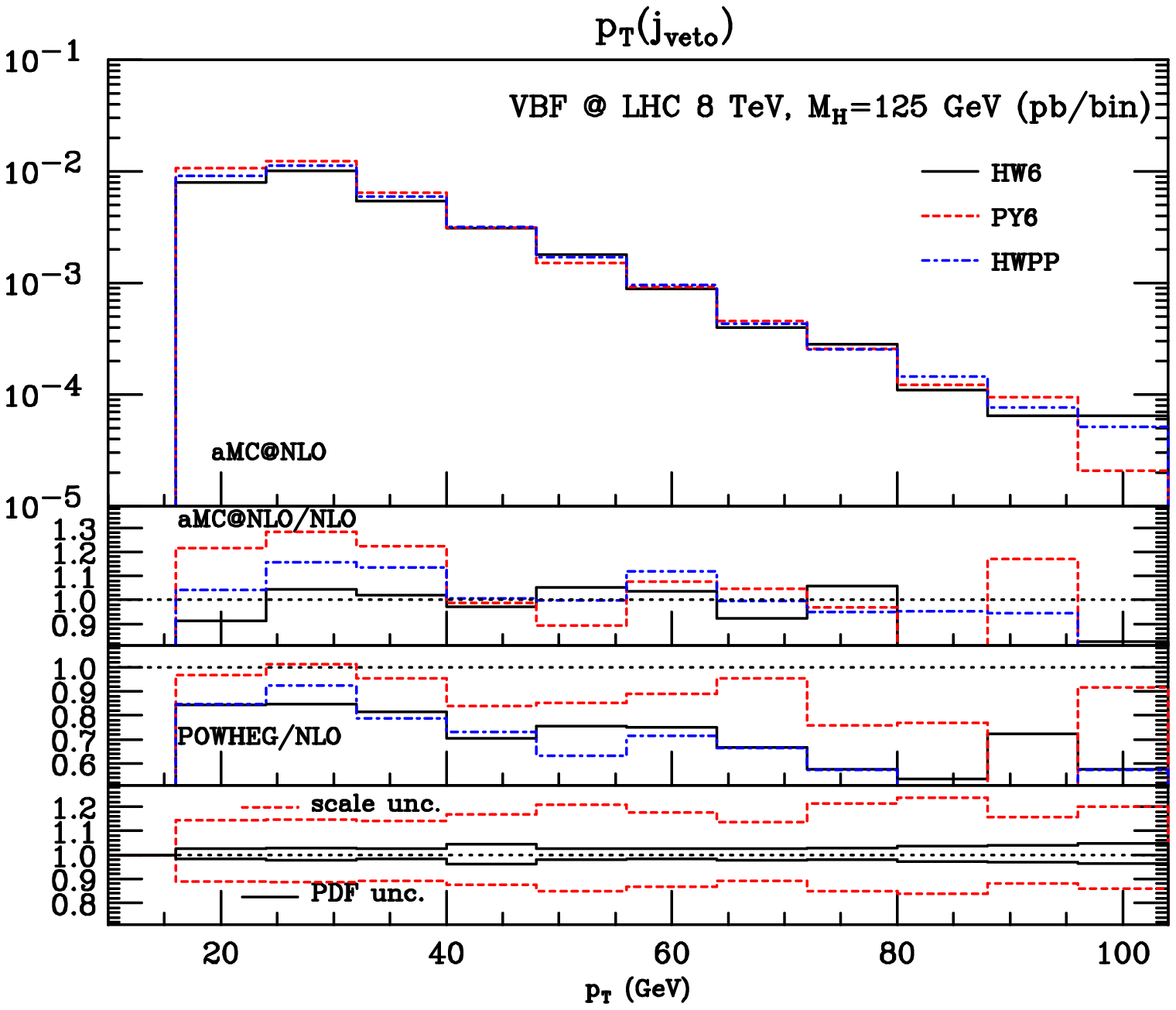}\\[5mm]
\includegraphics[width=0.75\textwidth]{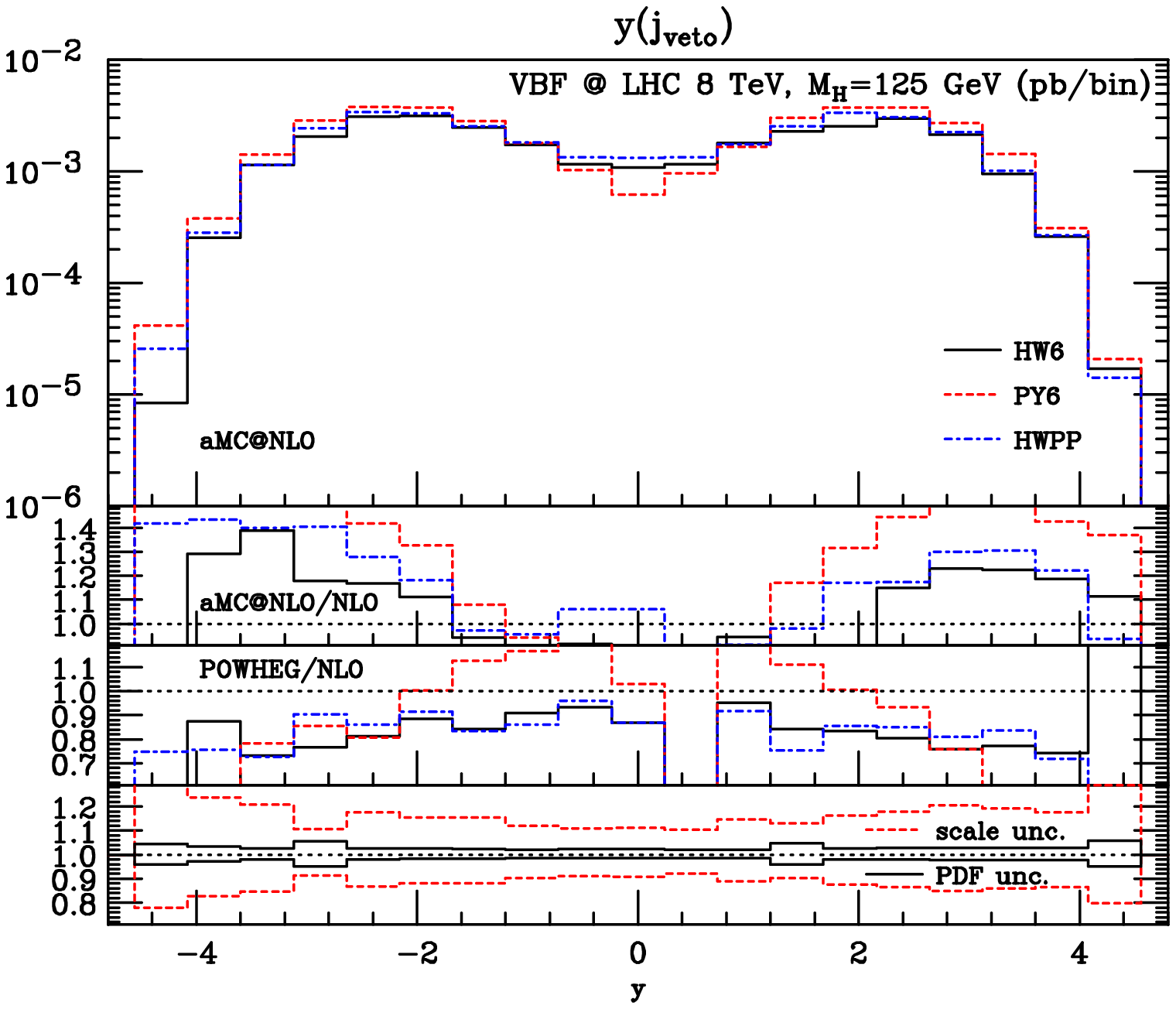}
\caption{\footnotesize
  \label{figamc:jveto}
    Same pattern as in figure \ref{figamc:higgs} for the veto-jet transverse momentum (top) and rapidity (bottom).}
\end{figure}

\section{Conclusions}
\label{secamc:concl}
We have presented a comparison between {\amcatnlo}, {\powheg}, and fixed-order-NLO predictions for VBF Higgs production at the 8 \tev~LHC. This study allows one to asses the NLO-matching systematics affecting this process and its various key-observables. Our results can be summarised as follows. For observables described with NLO accuracy at the parton level, predictions display small theoretical uncertainties, up to $\pm$5\%, and show a good agreement between the two matching schemes. Differences with respect to the pure-NLO predictions result from the action of the shower on the jet activity of the events. For quantities described with LO precision, theoretical uncertainties are consistently larger, of the order of $\pm$10\% to $\pm$15\%, as well as the discrepancy between the two matching prescriptions, with {\amcatnlo} generally closer to the fixed-order NLO than {\powheg}; in particular there is a visible effect in the observables related to the third jet. Still, all results are largely compatible once theoretical error bands are taken into account.

\section*{Acknowledgements}
We are grateful to Fabio Maltoni for many useful comments on the manuscript, and to Carlo Oleari for discussions. SF is on leave of absence from INFN, sezione di Genova. This work has been supported in part by the ERC grant 291377 ``LHCtheory: Theoretical predictions and analyses of LHC physics: advancing the precision frontier'', by the Forschungskredit der  Universit\"at Z\"urich, by the Swiss National Science Foundation (SNF) under contract 200020-138206 and by the Research Executive Agency (REA) of the European Union under the Grant Agreement number PITN-GA-2010-264564 (LHCPhenoNet). The work of MZ is supported by the IISN ``MadGraph'' convention 4.4511.10, the IISN ``Fundamental interactions'' convention 4.4517.08 and the Belgian IAP Program BELSPO P7/37.





\bibliographystyle{elsarticle-num}
\bibliography{vbfbib}







\end{document}

%% file: newcommand.tex
\newcommand{\gev}{\,\textrm{GeV}}

\newcommand{\pb}{\,\textrm{pb}}
\newcommand{\tev}{\,\textrm{TeV}}

\newcommand{\cf}[1]{{Fig.~\ref{#1}}}

\newcommand{\ct}[1]{{Tab.~\ref{#1}}}

\newcommand{\pythiasix}{\textsc{Pythia6}}

\newcommand{\herwigsix}{\textsc{HERWIG6}}
\newcommand{\herwigpp}{\textsc{HERWIG++}}

\newcommand{\powheg}{\textsc{POWHEG}}

\newcommand{\mcatnlo}{\textsc{MC@NLO}}
\newcommand{\madloop}{\textsc{MadLoop}}
\newcommand{\amcatnlo}{a\textsc{MC@NLO}}
\newcommand{\madfks}{\textsc{MadFKS}}
\newcommand{\madgraph}{\textsc{MadGraph}}

\newcommand{\fastjet}{\textsc{FastJet}}
\newcommand{\cuttools}{\textsc{CutTools}}

\newcommand{\captionamcatnlo}{Main frame: {\amcatnlo} matched with {\herwigsix} (black solid), virtuality-ordered {\pythiasix} (red dashed) and {\herwigpp} (blue dot-dashed). Upper (middle) inset: ratios of {\amcatnlo} ({\powheg}) over the fixed-order NLO, with the same colour pattern as the main frame. Lower inset: scale (red-dashed) and PDF (black solid) uncertainties for {\amcatnlo}+{\herwigsix}. See text for further details.}

\def\ie{{\it i.e.}}